\documentclass[cameraready]{Interspeech}
\usepackage{tabularx}
\usepackage{amsmath,graphicx,hyperref,caption,rotating}

\usepackage{booktabs}
\usepackage{multirow}
\usepackage{svg}
\usepackage{setspace}
\usepackage{todonotes}
\usepackage{cleveref}
\usepackage{microtype}
\usepackage{algorithm}
\usepackage{algpseudocode}
\usepackage{amssymb}
\hypersetup{colorlinks=true, linkcolor=blue, citecolor=green}

\title{AuDirector: A Self-Reflective Closed-Loop Framework for Immersive Audio Storytelling}

\author[affiliation={1}, equalcontribution]{Yiming}{Ren}
\author[affiliation={1}, equalcontribution]{Xuenan}{Xu}
\author[affiliation={1,2}]{Ziyang}{Zhang}
\author[affiliation={1}]{Wen}{Wu}
\author[affiliation={1}]{Baoxiang}{Li}
\author[affiliation={1,2},correspondingauthor]{Chao}{Zhang}

\address{
    $^1$ Shanghai Artificial Intelligence Laboratory \\
    $^2$ Tsinghua University
}

\email{ }


\usepackage{comment}

\begin{document}

\maketitle

\begin{abstract}
Despite advances in text and visual generation, creating coherent long-form audio narratives remains challenging.
Existing frameworks often exhibit limitations such as mismatched character settings with voice performance, insufficient self-correction mechanisms, and limited human interactivity.
To address these challenges, we propose AuDirector, a self-reflective closed-loop multi-agent framework.
Specifically, it involves an Identity-Aware Pre-production mechanism that transforms narrative texts into character profiles and utterance-level emotional instructions to retrieve suitable voice candidates and guide expressive speech synthesis, thereby promoting context-aligned voice adaptation.
To enhance quality, a Collaborative Synthesis and Correction module introduces a closed-loop self-correction mechanism to systematically audit and regenerate defective audio components.
Furthermore, a Human-Guided Interactive Refinement module facilitates user control by interpreting natural language feedback to interactively refine the underlying scripts.
Experiments demonstrate that AuDirector achieves superior performance compared to state-of-the-art baselines in structural coherence, emotional expressiveness, and acoustic fidelity. Audio samples can be found at \url{https://github.com/Riddae/AuDirector}.

\end{abstract}

\section{Introduction}
\label{intro}

Recent advances in generative modeling have yielded notable improvements in text generation~\cite{comanici2025gemini,achiam2023gpt} and visual synthesis, covering both images~\cite{wu2025qwen} and videos~\cite{brooks2024video}.
As an equally important modality, audio plays a critical role in multimedia content creation, and recent progress in generative models has correspondingly driven rapid developments in audio generation.
In the domain of text-to-speech (TTS), advanced systems can now perform high-fidelity voice cloning while following fine-grained emotional and prosodic~\cite{zhou2025indextts2,du2024cosyvoice}.
Meanwhile, text-to-audio (TTA) and text-to-music (TTM) models have shown strong capabilities in generating environmental sound effects~\cite{hung2024tangoflux,liu2023audioldm} and musical compositions that align closely with the input description~\cite{copet2023simple,agostinelli2023musiclm}.

Despite these advances, existing research in audio generation remains fragmented: most systems operate within a single domain (e.g., speech, music, or sound effects) and typically produce only short segments with limited structural coherence.
These limitations constrain the applicability of current audio generation models to more structured storytelling scenarios, which inherently require the harmonic integration of diverse sound types over extended durations.
This gap highlights the need for systems capable of holistic audio generation.

To address this gap, the research community has explored agent-based audio generation.
This paradigm was initially validated in other domains: ViperGPT~\cite{suris2023vipergpt} and HuggingGPT~\cite{shen2023hugginggpt} demonstrated that large language models (LLMs) can act as controllers to decompose and solve complex cross-modal tasks by generating executable programs or invoking domain expert models.
Inspired by this, works like AudioGPT~\cite{huang2024audiogpt} that integrate multi-task understanding have also emerged in the audio domain.
In audio storytelling, pioneering works such as WavJourney~\cite{liu2025wavjourney} and PodAgent~\cite{xiao2025podagent} leverage the reasoning capabilities of LLMs to generate executable scripts, constructing auditory narratives by integrating various audio foundational models.

Although agent-driven audio storytelling has advanced significantly, current frameworks still face several limitations:
First, constrained adaptive speech representation. They typically lack dynamic voice adaptation and fine-grained emotional control, leading to a mismatch between the generated speech and the story context. 
Second, the absence of a self-correcting quality control. By ignoring the inherent variability of generative models, these frameworks fail to detect and selectively regenerate low-quality clips, compromising the overall output quality. 
Third, limited human-in-the-loop collaboration. Operating predominantly in an open-loop manner, existing systems offer users little to no capability to refine specific audio elements during generation.

To address these challenges, we propose \textbf{AuDirector}, a multi-agent framework that automatically generates high-quality immersive audio through a closed-loop self-reflection mechanism comprising planning, execution, quality control, and feedback.
First, we address limited voice adaptation by introducing \textit{Identity-Aware Pre-production}. This mechanism extracts character profiles and emotional instructions from the narrative to direct voice retrieval and expression control, ensuring a seamless alignment between speech and story context.
Second, we establish a \textit{Collaborative Synthesis and Correction} module to enable closed-loop quality control. 
By auditing audio and triggering targeted regeneration for defects, this module significantly enhances output stability and fidelity.
Third, to mitigate the inflexibility of open-loop systems, we propose \textit{Human-Guided Interactive Refinement}.
By interactively refining the underlying script based on natural language user feedback, the system provides users with enhanced control over the final output.

The main contributions of this work are summarized as follows:
1) We propose AuDirector, a self-reflective closed-loop multi-agent framework facilitating coherent, emotionally expressive, and long-form audio storytelling across speech, sound effects, and background music.
2) We establish a collaborative quality control mechanism that integrates self-correction with human-computer interaction to enhance generation quality and support interactive control from users.
3) Experiments demonstrate that AuDirector achieves superior performance compared to state-of-the-art (SOTA) baselines in structural coherence, emotional expressiveness, and acoustic fidelity.

\begin{figure*}[htbp]
    \centering
    \includegraphics[width=\linewidth]{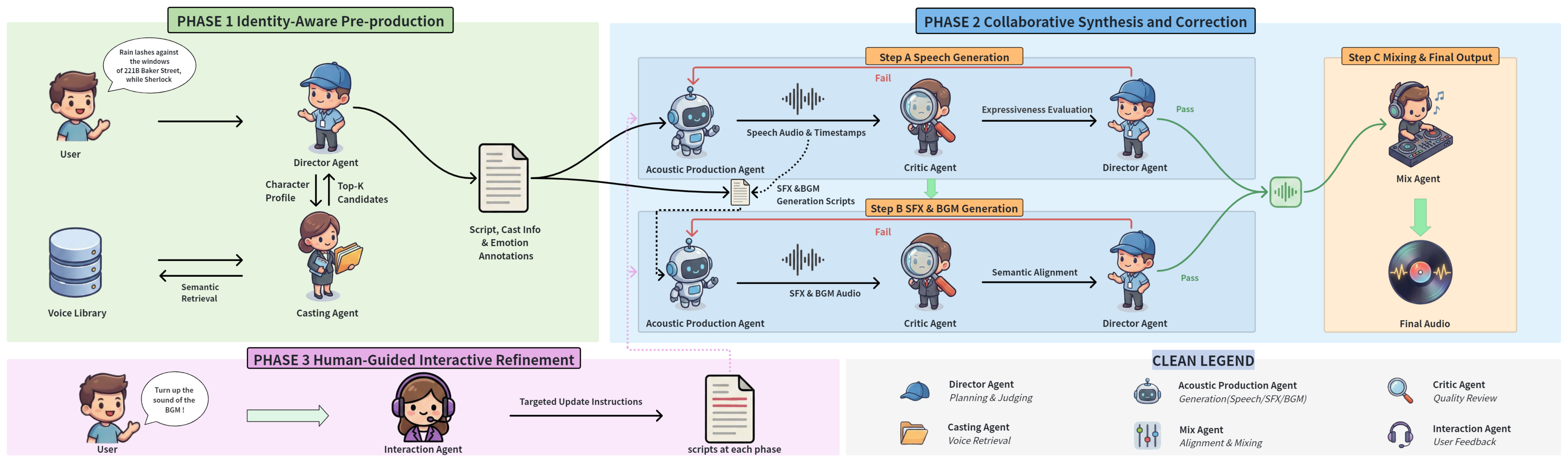}
    \caption{Overview of the AuDirector framework: 1) \textbf{Identity-aware pre-production} for script-driven voice casting; 2) \textbf{Collaborative synthesis and correction} featuring Critic-led quality auditing and self-correction; and 3) \textbf{Human-guided interactive refinement} for precise editing via natural language feedback.}
    \label{fig:AuDirector}
\end{figure*}

\section{AuDirector}
\label{sec:AuDirector}

\textbf{AuDirector} transforms user prompts $P_{user}$ (e.g., \textit{``Rain lashes against the windows of 221B Baker Street, while Sherlock Holmes reveals the truth with cold, analytical precision...''}) into high-fidelity audio narratives with rich sound effects and background music, through a collaborative multi-agent architecture (\Cref{fig:AuDirector}).
As formalized in \Cref{alg:audirector}, the generation pipeline proceeds through three core stages:

1) Identity-Aware Pre-production: 

The \textbf{Director Agent} ($\mathcal{A}_{dir}$) and \textbf{Casting Agent} ($\mathcal{A}_{cas}$) collaboratively orchestrate emotional script parsing and character casting.

2) Collaborative Synthesis and Correction: 
\textbf{Acoustic Production Agent }($\mathcal{A}_{aco}$) and Critic Agent ($\mathcal{A}_{cri}$) produce and refine multi-track audio through an iterative auditing loop. The resulting tracks are then integrated by the \textbf{Mix Agent} ($\mathcal{A}_{mix}$) to produce the initial output $A_{init}$.

3) Human-Guided Interactive Refinement: The \textbf{Interaction Agent} ($\mathcal{A}_{int}$) interprets natural language feedback to trigger targeted regeneration, enabling the Mix Agent to refine the final audio $A_{final}$.

\subsection{Identity-Aware Pre-production}
This stage begins with $\mathcal{A}_{dir}$, which leverages an LLM to transform the user prompt $P_{user}$ into a structured dialogue script $S_{dial}$ and character profiles $\{q_{id}\}$. This foundational process enables subsequent script-driven automated voice casting and dynamic emotion modeling.

\textbf{Voice Library Construction:} To support dynamic voice selection across varying narrative contexts, we construct an extensive and highly diversified voice library $\mathcal{D} = \{(a_i, d_i)\}_{i=1}^N$ comprising $N=320$ speech samples. This collection is curated to cover a broad spectrum of vocal identities, including diverse ages, genders, and a wide array of acoustic profiles (e.g., varying dialects and speaking styles).
For each audio sample $a_i$, a textual description $d_i$ is generated using Gemini-3-Pro and subsequently verified by manual inspection to ensure high acoustic quality and description accuracy.

\textbf{Runtime Voice Selection:} Voice selection follows a ``coarse-to-fine'' two-step retrieval process (Lines 2--4 in \Cref{alg:audirector}):
\begin{itemize}
    \item \textit{Semantic Filtering:} $\mathcal{A}_{cas}$ utilizes a text embedding module $E(\cdot)$ to map the character profile $q_{id}$ and candidate descriptions $d_i \in \mathcal{D}$ into a shared embedding space, filtering the top-$K$ candidates to form a high-recall candidate set $\mathcal{S}_{cand}$.
    \item \textit{Director Decision:} $\mathcal{A}_{dir}$ determines the optimal voice $a_{k^*}$ from $\mathcal{S}_{cand}$ by considering the dialogue script $S_{dial}$, ensuring that the selected vocal identity aligns with the emotional and character requirements of the script.
\end{itemize}

\textbf{Context-Driven Emotion Generation.} To enhance expressiveness and mitigate prosodic monotony, $\mathcal{A}_{dir}$ dynamically synthesizes an emotional profile for each dialogue utterance $T_{dial} \in S_{dial}$. 
Specifically, by evaluating the surrounding narrative scene context $C_{scene}$, $\mathcal{A}_{dir}$ translates the inferred affective intent into an explicit 7-dimensional instruction $I_{emo} \in \mathbb{R}^7$. 
Formally, $I_{emo}$ is represented as a weighted mixture over a predefined emotion basis $\mathcal{E}$:$$\mathcal{E} =
\left\{
\begin{aligned}
& \text{Anger, Happiness, Fear, Disgust,} \\
& \text{Sadness, Surprise, Neutral}
\end{aligned}
\right\}.$$

\subsection{Collaborative Synthesis and Correction}
This stage ensures high-fidelity audio quality through a nested ``generation-evaluation-refinement'' loop (lines 5--22 in \Cref{alg:audirector}):

\textbf{Hierarchical Synthesis.}
$\mathcal{A}_{aco}$ adopts a hierarchical synthesis pipeline. First, conditioned on $(T_{dial} \mid a_{k^*}, I_{emo})$, it generates the primary speech track $A_{speech}$. 
Subsequently, $\mathcal{A}{dir}$ uses the voice timestamp $t$ and the dialogue script $S{dial}$ to build a structured production script $S_{prod}$.
The script incorporates alignment information and ambient sound descriptions $d$.
Building upon this, it coordinates the generation of non-voice tracks $A_{ns}$ and provides the foundation for $\mathcal{A}{mix}$ to synthesize all tracks into the initial audio $A{init}$.

\textbf{Collaborative Correction.} To mitigate potential generative errors, the system employs a closed-loop refinement strategy:
\begin{itemize}
    \item \textbf{Speech Loop:} $\mathcal{A}_{cri}$ first generates an evaluative textual description of the synthesis quality of $A_{speech}$, and subsequently provides a corresponding quantitative score based on this assessment. If the score fails to meet the predefined threshold $\tau_{speech}$, $\mathcal{A}_{dir}$ refines the emotion instruction $I_{emo}$ to trigger regeneration.
    This process continues for up to $N_{max}$ attempts, with the system ultimately preserving the audio sample that achieves the highest quality score.
    \item \textbf{Non-speech Loop:}
    This loop follows the same iterative framework as the speech loop, but the score is evaluated as the semantic alignment between $A_{ns}$ and its textual description $d$. 
    If the score falls below the threshold $\tau_{ns}$, the textual prompt $d$ or the random seed is adjusted for regeneration. 
\end{itemize}

\begin{algorithm}[tb]
\caption{AuDirector Generation and Refinement}
\label{alg:audirector}
\setstretch{1.06} 
\begin{algorithmic}[1]
\Require $P_{user}$, $\mathcal{D}$, $\{\tau_{speech},\tau_{ns}\}$, $N_{max}$
\Ensure $A_{final}$

\Statex \hspace*{-1.4em} \textbf{I. Identity-Aware Pre-production}
\State $S_{dial}, \{I_{emo}\}, \{q_{id}\} \gets \mathcal{A}_{dir}(P_{user})$
\For{each $q_{id}$}
    \State $a_{k^*}\gets\mathcal{A}_{dir}\big(S_{dial}, \mathcal{A}_{cas}(q_{id},\mathcal{D},K)\big)$
\EndFor

\Statex \hspace*{-1.8em} \textbf{II. Collaborative Synthesis \& Correction}
\For{each $T_{dial}\in S_{dial}$}
    \For{$n=1 \dots N_{max}$}
        \State $A_{speech}^{(n)} \gets \mathcal{A}_{aco}(T_{dial}\mid a_{k^*},I_{emo})$
        \State $score^{(n)} \gets \mathcal{A}_{cri}(A_{speech}^{(n)})$
        \State \textbf{if} $score^{(n)} \ge \tau_{speech}$ \textbf{then break}
    \EndFor
    \State $A_{speech}\gets \arg\max_{n} score^{(n)}$
\EndFor
\State $S_{prod}\gets\mathcal{A}_{dir}(S_{dial},t_{speech})$

\For{each non-speech $(d,s)\in S_{prod}$}
    \For{$m=1 \dots N_{max}$}
        \State $A_{ns}^{(m)} \gets \mathcal{A}_{aco}(d,s)$
        \State $score^{(m)} \gets \mathcal{A}_{cri}(A_{ns}^{(m)},d)$
        \State \textbf{if} $score^{(m)} \ge \tau_{ns}$ \textbf{then break}
    \EndFor
    \State $A_{ns}\gets \arg\max_{m} score^{(m)}$
\EndFor
\State $A_{init}\gets\mathcal{A}_{mix}(\{A_{speech}\},\{A_{ns}\},S_{prod})$

\Statex \hspace*{-2.0em} \textbf{III. Human-Guided Interactive Refinement}
\State $A_{final}\gets A_{init}$
\While{$F_{user}$}
    \State $S_{upd}\gets\mathcal{A}_{int}(F_{user},S_{prod})$
    \State $A_{final}\gets
    \mathcal{A}_{mix}(TR(S_{upd}), 
    A_{final})$
\EndWhile

\State \Return $A_{final}$
\end{algorithmic}
\end{algorithm}

\subsection{Human-Guided Interactive Refinement}
\label{sec:Human}
 
To facilitate direct human intervention, $\mathcal{A}_{int}$ serves as a human–computer interface by leveraging the reasoning and semantic parsing capabilities of an LLM (\Cref{alg:audirector}, Lines 23--27).
Upon receiving natural language feedback $F_{user}$ (e.g., ``lower the background music volume''), $\mathcal{A}_{int}$ interprets the intent and precisely modifies the corresponding attributes within the structured production script $S_{prod}$. 
Only affected components (e.g., a specific sound effect) are regenerated, denoted as Targeted Regeneration (TR).
$\mathcal{A}_{mix}$ then updates the mixed track $A_{final}$ based on the regenerated components.
This significantly reducing computational costs during interactive editing.

\begin{table*}[t]
\centering
\caption{Comparison of different methods on Objective and Subjective Metrics. Objective (AES): Production Quality (PQ), Production Complexity (PC), Content Enjoyment (CE), Content Usefulness (CU), and Voice-Role Matching (VRM). Subjective (MOS):  Matching (M), Quality (Q), Alignment (Ali), Emotion (Emo), Aesthetic (Aes).}
\label{tab:metrics_wide}
\renewcommand{\arraystretch}{1.2}
\setlength{\tabcolsep}{0pt}

\begin{tabular*}{\textwidth}{@{\extracolsep{\fill}} l ccccc ccccc }
\toprule
\multirow{2}{*}{\textbf{Method}} & \multicolumn{5}{c}{\textbf{Objective Metrics}} & \multicolumn{5}{c}{\textbf{Subjective Metrics}} \\
\cmidrule(lr){2-6} \cmidrule(lr){7-11}
& CE~$\uparrow$ & CU~$\uparrow$ & PC~$\uparrow$ & PQ~$\uparrow$ & VRM~$\uparrow$ & MOS-M~$\uparrow$ & MOS-Q~$\uparrow$ & MOS-Ali~$\uparrow$ & MOS-Emo~$\uparrow$ & MOS-Aes~$\uparrow$ \\
\midrule
WavJourney              & 5.19 & 5.66 & \textbf{4.42} & 6.95 & 2.61 & 3.09 $\pm$ 0.67 & 3.58 $\pm$ 0.45 & 3.30 $\pm$ 0.61  & 3.10 $\pm$ 0.52  & 3.41 $\pm$ 0.62  \\
PodAgent                & 6.37 & \textbf{7.11} & 2.98 & 7.46 & 3.59 & 3.48 $\pm$ 0.59 & 3.73 $\pm$ 0.47 & 3.60 $\pm$ 0.54 & 3.60 $\pm$ 0.50 & \textbf{4.04 $\pm$ 0.45}  \\
AuDirector (w/o Critic) & 6.22 & 6.52 & 4.18 & 7.37 & 4.23 & \textbf{4.01 $\pm$ 0.34} & 3.83 $\pm$ 0.44 & 3.65 $\pm$ 0.50 & 4.00 $\pm$ 0.37  & 3.92 $\pm$ 0.46 \\
AuDirector (Full)       & \textbf{6.46} & 6.98 & 4.32 & \textbf{7.59} & \textbf{4.23} & 4.00 $\pm$ 0.32 & \textbf{3.86 $\pm$ 0.42} & \textbf{3.74 $\pm$ 0.44}  & \textbf{4.17 $\pm$ 0.45}  & 4.01 $\pm$ 0.38 \\
\bottomrule
\end{tabular*}
\end{table*}

\section{Experimental Setups}

\subsection{Implementation Details}

AuDirector is a collaborative multi-agent system with Gemini-3-Pro serving as the Director and Interaction Agents. We utilize EmbeddingGemma~\cite{vera2025embeddinggemma} for casting and employ IndexTTS2~\cite{zhou2025indextts2}, TangoFlux~\cite{hung2024tangoflux}, and MusicGen~\cite{copet2023simple} for speech, SFX, and BGM production, respectively. The Critic Agent leverages MIMO-Audio~\cite{coreteam2025mimoaudio} and CLAP~\cite{wu2023large} for quality assessment, followed by \texttt{pydub} for final composition. Detailed system prompts will be available in our open-source repository.

\subsection{Evaluation Data}
\label{sec:benchmark} 
Our evaluation dataset comprises 100 diverse scenarios categorized into two primary genres: Podcasts (40 topics) and Radio Dramas (60 stories).

\textbf{Podcasts:} We select a subset from Vicuna \cite{chiang2023vicuna}, focusing on four categories: \textit{Generic}, \textit{Knowledge}, \textit{Common-sense}, and \textit{Counterfactual}. Each category contains 10 topics aimed at evaluating multi-turn conversational depth and reasoning. 

\textbf{Radio Dramas:} We use 60 narratives from ROCStories \cite{mostafazadeh2016corpus}, leveraging their strong causal-temporal structure to evaluate the model’s ability to expand five-sentence premises into complex, multi-character scripts with rich audio design.

\subsection{Evaluation Protocols}

\subsubsection{Overall Generation Quality} 
We evaluate the performance of AuDirector by comparing it with two baseline systems: \textbf{WavJourney} and \textbf{PodAgent}. To ensure a fair comparison of orchestration capabilities, all systems are driven by the same LLM and utilize identical underlying audio generation backends as AuDirector.
We also incorporate \textbf{AuDirector (w/o Critic)} to validate the effectiveness of the self-correction mechanism.
The evaluation is conducted through both objective and subjective metrics.

\textbf{Objective Evaluation:} We employ automated scoring using LLMs and pre-trained evaluation models.
\textbf{Voice-Role Matching (VRM)}: We leverage Gemini-3-Pro to assess the alignment between the selected voice and the Character Profile generated by $\mathcal{A}_{dir}$. The model assigns a score on a 1–-5 scale (1: poor match; 5: perfect match) based on key dimensions, including gender, age, timbre, and delivery style.
\textbf{Audio Aesthetics Score (AES)}~\cite{tjandra2025meta}: This metric provides scores across four aspects: \textit{Production Quality (PQ)} for technical fidelity; \textit{Production Complexity (PC)} for acoustic layering richness; \textit{Content Enjoyment (CE)} for artistic appeal; and \textit{Content Usefulness (CU)} for practical utility.

\textbf{Subjective Evaluation:} Complementarily, a subjective Mean Opinion Score (MOS) study involving 10 evaluators assesses perceptual performance. Evaluators provide scores on a 1--5 scale across five dimensions:
\textit{MOS-Matching (M)} for the consistency between the voice performance and character settings within the story context;
\textit{MOS-Quality (Q)} for general audio clarity and quality;
\textit{MOS-Alignment (Ali)} for the semantic and temporal alignment of SFX and BGM with the text;
\textit{MOS-Emotion (Emo)} for prosodic vividness and emotional expressiveness;
\textit{MOS-Aesthetic (Aes)} for overall artistic appeal and listening comfort.

\subsubsection{Interaction Evaluation}
Since human-guided interactive refinement is achieved by modifying corresponding scripts, we evaluate the script modification accuracy.
We curate a set of 200 natural language editing instructions based on \Cref{sec:benchmark} for evaluation.
These instructions are evenly distributed across four key dimensions: 
\begin{itemize}
\item 1) \textit{Speech Refinement}: adjusting vocal attributes or spoken content, e.g., ``make the tone more sorrowful'';  
\item 2) \textit{Acoustic Content Modification}: altering non-speech elements, e.g., ``change the rain sound to a storm'';  
\item 3) \textit{Signal Gain Control}: adjusting volume levels, e.g., ``lower the background music volume'';  
\item 4) \textit{Structural Editing}: inserting or deleting audio segments, e.g., ``insert a scream here''.
\end{itemize}

Performance is assessed via \textbf{Instruction Execution Accuracy (IEA)} , where each modification is manually verified for alignment with the user's intent.

\section{Results and Analysis}

\subsection{Analysis of Overall Generation Quality}
\textbf{Objective Evaluation}
As shown in \Cref{tab:metrics_wide}, AuDirector leads in PQ, CE, and VRM. The VRM advantage confirms the effectiveness of the Casting Agent in achieving precise voice selection via a ``coarse-to-fine'' retrieval process.
In contrast, baselines rely on coarse metadata and exhaustive prompt-based selection by the LLM, which degrades both precision and scalability as the audio library expands.

Regarding the performance in PC and CU, while WavJourney achieves high PC through diverse audio elements, its poor coordination results in the lowest scores elsewhere.
Conversely, PodAgent's fixed format yields high CU at the expense of complexity.
Notably, AuDirector ranks second in both dimensions, striking an optimal balance between acoustic richness and practical utility through the harmonious integration of speech and environmental tracks.

\textbf{Subjective Evaluation.} 
The subjective evaluations shown on the right side of \Cref{tab:metrics_wide} further validate the superiority of AuDirector.
The best performance in \textit{MOS-M} reflects users' effectiveness of AuDirector in casting and character-role alignment.
The significant improvement in \textit{MOS-Emo} can be attributed to the Director Agent's dynamic emotional instructions, which effectively enhance vocal expressiveness.
In addition, the relatively high \textit{MOS-Ali} and \textit{MOS-Aes} scores highlight the semantic alignment and coherence between different audio components (speech, BGM, and SFX).
Notably, since all methods employ a unified acoustic production backend, the differences in \textit{MOS-Q} are not significant, thereby ensuring the fairness of the comparison at the agent scheduling level.

\textbf{Ablation Study}
The comparison between AuDirector and its \textit{w/o Critic} variant further validates the effectiveness of the closed-loop self-correction mechanism. 
Specifically, notable gains are observed across all metrics except \textit{MOS-Q} and \textit{MOS-M}. 
These improvements demonstrate the effectiveness of the Critic Agent in alleviating inherent generation quality variability in $\mathcal{A}_{aco}$ to improve the final audio quality.

\begin{table}[t]
    \centering
    \caption{Evaluation of Human-Guided Interactive Refinement. We report the Instruction Execution Accuracy (IEA) across four categories.}
    \label{tab:interaction_results}
    \begin{tabular}{lc} 
        \toprule
        \textbf{Instruction Category} & \textbf{IEA (\%)} \\
        \midrule
        Signal Gain Control           & 96.00 \\
        Structural Editing            & 84.00 \\
        Speech Refinement             & 92.00 \\
        Acoustic Content Modification & 88.00 \\
        \midrule 
        Overall Average               & \textbf{90.00} \\
        \bottomrule
    \end{tabular}
\end{table}

\subsection{Interaction Evaluation}
As presented in \Cref{tab:interaction_results}, the interactive editing achieves an overall IEA of 90.00\%.
Signal Gain Control and Speech Refinement demonstrate high accuracy.
This is largely due to simple and explicit positional cues, such as sequential numbers (e.g., ``the 3rd dialogue'', ``the 2nd BGM''), which help clearly identify the target.
In contrast, Acoustic Content Modification and Structural Editing exhibit a slight performance drop.
These tasks necessitate more complex and fine-grained temporal localization, particularly in dense acoustic segments with overlapping SFX.
For instance, distinguishing a specific target within a scene containing overlapping elements (e.g., footsteps mixed with rain) leads to ambiguity for the agent, thereby increasing the difficulty of precise script editing.

\section{Conclusion}

This paper presents AuDirector, a multi-agent framework designed to enhance immersive audio storytelling through closed-loop collaboration. 
By integrating identity-aware pre-production with a self-reflective quality control loop, the framework ensures that generated audio maintains high-level semantic consistency with the narrative while granting users fine-grained control over the final output. 
Experimental results confirm that AuDirector significantly improves character-role matching and emotional expressiveness.
However, we observed that current generative models still have limitations in generating non-speech audio tracks, especially in terms of acoustic diversity and nuance (e.g., smooth breathing versus tense breathing), which can lead to auditory inconsistencies and thus disrupt immersion. 
Future work will focus on fine-grained modeling of environmental sounds to bolster narrative consistency.

\section{Generative AI Use Disclosure}
In this work, generative AI was exclusively utilized to fix grammatical mistakes and adjust terminology, while all core research activities, including study design, data collection, analysis, and scientific reasoning, were conducted independently by the authors.

\bibliographystyle{IEEEtran}
\bibliography{mybib}

\end{document}